\documentclass[lettersize,journal]{IEEEtran}
\usepackage{cite}
\usepackage{amsmath,amssymb,amsfonts}
\usepackage{algorithmic}
\usepackage{graphicx}
\usepackage{textcomp}
\usepackage{amsmath}
\usepackage{amssymb}
\usepackage{latexsym}
\usepackage{CJK}
\usepackage{booktabs}
\usepackage{array}
\usepackage{multirow}
\usepackage{graphicx}
\usepackage{epstopdf}
\usepackage{caption}

\usepackage{algorithm}
\usepackage{algorithmic}
\usepackage{tabularx}
\usepackage{stfloats}
\usepackage{xcolor}
\usepackage{booktabs}

\usepackage{makecell}
\usepackage{threeparttable}
\usepackage{bm}
\usepackage{subfig}
\usepackage{float}

\usepackage{bbding}

\ifCLASSINFOpdf
\else
\fi

\begin{document}
%
\title{Cooperative Beam Selection for RIS-Aided Terahertz MIMO Networks via Multi-Task Learning}


\author{
       Xinying Ma, Gong Chen, and Xiaofei Wang
\thanks{
X.Y. Ma, G. Chen and X. F. Wang are with the Chengdu Innovation Center for Intelligent Unmanned Aerial System, Chengdu, China (e-mails: xymarger@126.com, 2337785477@qq.com, 1148993303@qq.com).

}
}


\maketitle

\begin{abstract}
Reconfigurable intelligent surface (RIS) have been cast as a promising alternative to alleviate blockage vulnerability and enhance coverage capability for terahertz (THz) communications. Owing to large-scale array elements at transceivers and RIS, the codebook based beamforming can be utilized in a computationally efficient manner. However, the codeword selection for analog beamforming is an intractable combinatorial optimization (CO) problem. To this end, by taking the CO problem as a classification problem, a multi-task learning based analog beam selection (MTL-ABS) framework is developed to implement cooperative beam selection concurrently at transceivers and RIS. In addition, residual network and self-attention mechanism are used to combat the network degradation and mine intrinsic THz channel features. Finally, the network convergence is analyzed from a blockwise perspective, and numerical results demonstrate that the MTL-ABS framework greatly decreases the beam selection overhead and achieves near optimal sum-rate compared with heuristic search based counterparts.
\end{abstract}



\section{Introduction}
With the ongoing development of sixth generation (6G),  the Internet of Things (IoT) has become an unprecedented proliferation, aiming at realizing diverse application scenarios and improving the quality of our lives\cite{introduction_01}. To support these bandwidth-hungry IoT applications, developing new spectrum resources proves to be an indispensable way \cite{introduction_02}. Intriguingly, terahertz (THz) band has been regarded as a promising solution for 6G wireless networks to provide sufficient spectrum resources \cite{introduction_03}. However, THz communication inevitably faces some intrinsic deficiencies when it is used in practice. On the one hand, THz waves experience serious propagation attenuation, leading to the transmission distance of THz communication within a small area. On the other hand, THz signals endure poor diffraction owing to the short wavelength \cite{introduction_04}. On account of these reasons, THz communications are sensitive to different kinds of obstacles \cite{introduction_05}, which results in coverage vulnerability and communication deterioration.

To tackle these challenges faced by THz communications, one of the most promising solutions is the massive multiple-input multiple-output (MIMO) technology. Thanks to the extremely small wavelength of THz waves, a large amount of array elements can be packaged within a limited physical area. Due to this factor, hybrid beamforming based on massive MIMO can be leveraged to compensate for the severe path loss \cite{introduction_06, introduction_07}. Nevertheless, the utilization of massive MIMO in THz communications also generates the narrow beams and thus is sensitive to the obstacles, which easily results in communication outages. Intriguingly, a promising technology called reconfigurable intelligent surface (RIS) is newly proposed to mitigate blockage vulnerability and improve coverage capability for THz wireless networks \cite{introduction_08,introduction_09,introduction_10,introduction_10a}. Specifically, the RIS belongs to the physical meta-surface family and is composed of a large number of passive reflecting elements. The RIS is commonly considered as a low-cost and energy-efficient paradigm shift for THz communications since it is designed without active devices \cite{introduction_11}. In addition, each reflecting element of RIS can passively manipulate the propagation direction of the impinging THz waves by adjusting the phase shifts and reflecting amplitudes. The passive RIS is capable of regulating the wireless channels and combating the undesired propagation conditions \cite{introduction_12}. With the dense deployment of RISs, the cunning corners can be well covered by THz signals. Benefited from these features, RISs can be widely applied to future THz communication scenarios.

\subsection{Prior Works}
As described above, taking advantage of the combination of massive MIMO and RIS for THz communications is an indispensable trend to enhance the network capacity and to reduce the energy consumption \cite{introduction_13}. To achieve this, one of the most critical solutions is to apply the beamforming technique to the THz system with massive array elements. However, differing from conventional systems without RIS, the RIS-aided THz system requires to jointly optimize the passive beamforming at the RIS and the active beamforming at the transceivers \cite{introduction_14,introduction_15,introduction_16}. To be specific, the work in \cite{introduction_15} initially proposes the concept of jointly passive and active beamforming design for RIS-aided multiple-input single-output (MISO) wireless network, and both semidefinite relaxation (SDR) and alternating optimization methods are employed to get a high-quality approximate solution as well as a lower bound. In terms of the point-to-point RIS-assisted MIMO system, the authors in \cite{introduction_16} consider the spectral efficiency maximization problem by jointly optimizing the reflection coefficients, the precoder and the combiner. In addition, unlike the works that use continuous phase shifts of RIS, the authors in \cite{introduction_17} propose a pragmatic RIS model with a limited number of discrete phase shifts to meet the requirement of practical applications. Then the sum-rate maximization problem is investigated by solving digital beamforming at the BS and discrete analog beamforming at the RIS. It is worth noting that most of the reported works pay more attention to designing active and passive beamforming through conventional optimization methods \cite{introduction_18,introduction_19,introduction_20}, which are more likely to be restricted by the complex and non-convex issues.

In addition, both active MIMO and passive RIS possess an extremely large number of array elements at THz band, but current research works tend to optimize the phase shifts of RIS elements one by one during the signal processing stage, which definitely leads to high latency and heavy computational complexity for RIS-aided THz communication systems \cite{introduction_30}. Hence, it is more practical for massive MIMO and RIS techniques to adopt the subarray architecture and each radio frequency (RF) chain only connects to a subset of antennas \cite{introduction_31}. Given the subarray structure, selecting analog beams for transceiver and RIS is an efficient way to guarantee the high sum-rate performance \cite{introduction_35}. It should be noted that conventional MIMO systems only carry out the beam selection procedure at the transceiver sides \cite{introduction_36,introduction_37}. By contrast, the RIS-aided THz multi-user MIMO (MU-MIMO) system requires to achieve analog beam selection at the base station (BS), the RIS and users, respectively. In other words, there are three different kinds of beam selection targets that need to be considered simultaneously. To cope with this complex combinatorial optimization (CO) problem, the heuristic beam search algorithms are time-consuming and computationally infeasible in real-time communications.
Fortunately, deep learning (DL) can be cast as a promising technology to address high-dimension, complex environment, non-linear, non-convex and other mathematically intractable problems of wireless networks \cite{introduction_21,introduction_22,introduction_23}. More specifically, the DL that belongs to a subclass of machine learning can be used to obtain near-optimal performances for RIS-enabled communication systems by establishing a mapping relationship between channel state information (CSI) and objective function, e.g., sum-rate, energy efficiency, optimal precoders.

\subsection{Our Contributions}
Motivated by the mathematically intricate CO problem, a multi-task learning based analog beam selection (namely MTL-ABS) framework is developed to solve the beam selection problem in a low-complexity way for the RIS-enabled THz MU-MIMO systems. The MTL technique is a promising paradigm in machine learning communities and aims to exploit the useful information in the multiple related tasks to improve the generalization performance of all the tasks with low complexity \cite{introduction_38,introduction_39,introduction_40}. With more training data, MTL is able to learn more robust and general representations for multiple tasks, obtaining better performance and lower overfitting risk of each task. Benefiting from the advantages of MTL, the major goal of this paper is how to formulate the beam selection problem with inter-related tasks as a MTL problem. The main contributions can be summarized as follows.

\begin{itemize}
  \item   We primitively formulate a codebook-based beam selection problem for the RIS-enabled THz MU-MIMO system, where the subarray architecture is employed at the BS and RIS, respectively. In light of this system model, we derive a novel sum-rate metric to measure the beam selection performance.
  \item  The MTL-ABS framework is developed to tackle the complex CO problem in a low-complexity manner, which cooperatively solves multiple beam classification tasks by mining the features of data samples. More importantly, the MTL-ABS model shares same channel representations in the shared layer and then is divided into different task-specific layers to handle the beam selection problem in parallel.
  \item   Residual network (ResNet) is utilized to ease the substantially deeper network and overcome the network degradation, which ensures much higher accuracy from considerably increased network depth. To further extract the ResNet, we convert the two-dimension (2D) channel into three-dimension (3D) channel that deeply couples with the convolution layers existing in the ResNet.
  \item   Self-attention mechanism, which is defined as mapping a query and a set of key-value pairs to an output, is used in MTL-ABS model to improve the beam selection accuracy at the BS. With the help of the self-attention , the MTL-ABS model realizes different weight assignments for the same input channel samples, so as to obtain different types of importance for diverse subarrays.
  \item   The convergence of our proposed MTL-ABS network is analyzed from a blockwise perspective and the MTL-ABS framework is evaluated by training loss and validation accuracy. Moreover, numerical results demonstrate that the MTL-ABS framework can realize a better compromise between sum-rate performance and complexity compared with heuristic search algorithms.
\end{itemize}

The remainder of this paper is organized as follows. The RIS-enabled THz MIMO system model and beam selection problem are respectively introduced in Section II. Section III presents the proposed MTL-ABS framework. Section IV provides the network convergence and computational complexity. Finally, simulation results and conclusion are presented in Section V and Section VI, respectively.

\section{System and Channel Model}
In this section, we introduce the RIS-enabled THz MU-MIMO system model and channel model. Based on these assumptions, the beam selection problem is formulated.

\subsection{System Model}
%

We consider a hybrid beamforming architecture for the uplink RIS-assisted THz MU-MIMO system, where a BS employs $N_r$ antennas and $N_s$ RF chains to serve $K$ users with a single RF chain that connects to $N_t$ antennas. Each RF chain at the BS connects to ${N_{\rm{B}}}$ antennas, and thus we have ${N_r} = {{N_{\rm{B}}}} \times {N_s}$. Due to the poor diffraction of THz waves, the line-of-sight (LoS) path is usually blocked by the obstacles, and thus the RIS equipped with $M$ passive elements is utilized to assist the non-line-of-sight (NLoS) paths. Note that there is a controller between the BS and the RIS realizing the phase adjustment. We divide the RIS into $M_s$ virtual subarray blocks, and each subarray has $M_{\rm{B}}$ passive reflecting elements. To exploit the multiplexing gain, we assume a constraint condition of $K \le {N_s}$.

To begin with, we suppose that each user transmits a dedicated data stream to the BS. Thus, the transmitted signal for $k$th  user through the analog beam can be written as
\begin{align}\label{system01}
{{\bf{x}}_k} = {{\bf{f}}_k}{s_k},
\end{align}
where ${{\bf{f}}_k} \in {{\mathbb C}^{{N_t} \times 1}}$  is the analog beam for $k$th  user, and  ${s_k} \in {{\mathbb C}^{1 \times 1}}$ is the transmitted data symbol of $k$th user, respectively. For simplification, let ${\bf{F}} = \left[ {{{\bf{f}}_1},{{\bf{f}}_2}, \cdots ,{{\bf{f}}_2}} \right]$  denote the beam set of all $K$ users. The transmit power for $k$th user can be defined as ${\mathbb E} [ {|| {{{\bf{x}}_k}} ||_F^2} ] \le P$, where $P$ represents the maximum power for each user.

Considering the received signals for all $K$ users, the received signal ${\bf{y}} \in {{\mathbb C}^{{N_r} \times 1}}$ at the BS can be written as
\begin{align}\label{system02}
{\bf{y}} = {{\bf{{H}}}_r}{\bf{\Phi }}\sum\limits_{k = 1}^K {{{\bf{{H}}}_k}{{\bf{f}}_k}{s_k}}  + {\bf{n}},
\end{align}
where ${{\bf{H}}_r} \in {{\mathbb C}^{{N_r} \times M}}$  is the reflected BS-RIS channel, ${{\bf{H}}_k} \in {{\mathbb C}^{M \times {N_t}}}$ is the RIS-user channel of $k$th user, and ${\bf{n}} \in {{\mathbb C}^{{N_r} \times 1}}$ represents the additive white Gaussian noise with ${\cal C}{\cal N}\left( {0,{N_0}{{\bf{I}}_{{N_r}}}} \right)$. In addition, ${\bf{\Phi }} \in {{\mathbb C}^{M \times M}}$ is the diagonal phase shift matrix of the RIS, which can be represented as
\begin{align}\label{system03}
{\bf{\Phi }} = {\rm {diag}}\left( {{\varphi _{1,1}}, \cdots ,{\varphi _{m,b}}, \cdots ,{\varphi _{{M_s},{M_{\rm{B}}}}}} \right)
\end{align}
where each diagonal entry of ${\bf{\Phi }}$  is defined as ${\varphi _{m,b}} \buildrel \Delta \over = {e^{j{\theta _{m,b}}}}$ $\left( {\forall m \in [1,{M_s}], \forall b \in [1,{M_{\rm{B}}}]} \right)$ and the phase shift vector for $m$th subarray of the RIS can be further defined as ${{\bf{\Lambda }}_m} \buildrel \Delta \over = \left( {{\varphi _{m,1}},{\varphi _{m,2}}, \cdots ,{\varphi _{m,{M_{\rm{B}}}}}} \right) \in {{\mathbb C}^{1 \times {M_{\rm{B}}}}}$ , respectively.

Then, the BS selects analog beams to process the received signals in the analog domain, the received uplink signal can be formulated as
\begin{align}\label{system04}
{{\bf{y}}_{\rm{R}}} &= {{\bf{W}}_{\rm{R}}}{{\bf{{H}}}_r}{\bf{\Phi }}\sum\limits_{k = 1}^K {{{\bf{{H}}}_k}{{\bf{f}}_k}{s_k}}  + {{\bf{W}}_{\rm{R}}}{\bf{n}} \notag \\
& = \left[ {{{\bf{W}}_{\rm{R}}}{{\bf{{H}}}_r}{\bf{\Phi }}{{\bf{{H}}}_1}{{\bf{f}}_1}, \cdots ,{{\bf{W}}_{\rm{R}}}{{\bf{{H}}}_r}{\bf{\Phi }}{{\bf{{H}}}_K}{{\bf{f}}_K}} \right]\left[ {\begin{array}{*{20}{c}}
{{s_1}}\\
 \vdots \\
{{s_K}}
\end{array}} \right] + {{\bf{W}}_{\rm{R}}}{\bf{n}} \notag \\
& = \left[ {{{{\bf{\bar h}}}_1},{{{\bf{\bar h}}}_2}, \cdots ,{{{\bf{\bar h}}}_K}} \right]\left[ {\begin{array}{*{20}{c}}
{{s_1}}\\
 \vdots \\
{{s_K}}
\end{array}} \right] + {{\bf{W}}_{\rm{R}}}{\bf{n}} \notag \\
& = {\bf{\bar {H}s}} + {{\bf{W}}_{\rm{R}}}{\bf{n}},
\end{align}
where ${\bf{s}} = {\left[ {{s_1}, \cdots ,{s_K}} \right]^T}$, ${{\bf{\bar h}}_k} = {{\bf{W}}_{\rm{R}}}{{\bf{{H}}}_r}{\bf{\Phi }}{{\bf{{H}}}_k}{{\bf{f}}_k}$ and ${\bf{\bar {H}}} = \left[ {{{{\bf{\bar h}}}_1},{{{\bf{\bar h}}}_2}, \cdots ,{{{\bf{\bar h}}}_K}} \right]$, respectively. Due to the subarray architecture at the BS, ${{\bf{W}}_{\rm{R}}} \in {{\mathbb C}^{{N_s} \times {N_r}}}$ is a block diagonal combining matrix, which can be described as
\begin{align}\label{system05}
{{\bf{W}}_{\rm{R}}} = \left[ {\begin{array}{*{20}{c}}
{{{\bf{w}}_1}}& \cdots &0\\
 \vdots & \ddots & \vdots \\
0& \cdots &{{{\bf{w}}_{{N_s}}}}
\end{array}} \right],
\end{align}
where ${{\bf{w}}_n} \in {{\mathbb C}^{1 \times {N_{\rm{B}}}}}$ denotes the analog beam for $n$th subarray block at the BS.

In the digital beamforming domain, the zero-force (ZF) technique is considered to detect the received signals for $K$ users. Hence, the digital combining matrix can be written as
\begin{align}\label{system06}
{{\bf{W}}_{\rm{B}}} = {\left( {{{{\bf{\bar {H}}}}^H}{\bf{\bar {H}}}} \right)^{ - 1}}{{\bf{\bar {H}}}^H}.
\end{align}

After being processed by the digital combiner, the received signal can be further expressed as
\begin{equation}\label{system07}
\begin{aligned}
{{\bf{y}}_{\rm{B}}} &= {{\bf{W}}_{\rm{B}}}\left( {{\bf{\bar {H}s}} + {{\bf{W}}_{\rm{R}}}{\bf{n}}} \right) \\
&= {\bf{s}} + {\left( {{{{\bf{\bar {H}}}}^H}{\bf{\bar {H}}}} \right)^{ - 1}}{{{\bf{\bar {H}}}}^H}{{\bf{W}}_{\rm{R}}}{\bf{n}}.
\end{aligned}
\end{equation}

\subsection{Channel Model}
The cascaded channel for the RIS-enabled THz MU-MIMO system is comprised of BS-RIS channel ${{\bf{H}}_r}$, phase shift matrix ${\bf{\Phi }}$ and RIS-user channel  ${{\bf{H}}_k}\left( {\forall k = 1, \cdots ,K} \right)$. Considering the sparse nature of THz channel, we employ the widely applicable geometric Saleh-Valenzuela channel model with limited scattering paths \cite{system_01}. Specifically, the BS-RIS channel ${{\bf{H}}_r}$ can be given by
\begin{align}\label{system08}
{{\bf{H}}_r} = \sqrt {\frac{{{N_r}M}}{{{L_{\rm{B}}} + 1}}} \sum\limits_{l = 0}^{{L_{\rm{B}}}} {{\alpha _l}{{\bf{a}}_{\rm{B}}}\left( {\phi _l^{\rm{B}},\upsilon _l^{\rm{B}}} \right)} {\bf{a}}_{\rm{R}}^H\left( {\phi _l^{\rm{R}},\upsilon _l^{\rm{R}}} \right),
\end{align}
where ${L_{\rm{B}}}$ is the number of NLoS paths between the BS and the RIS and ${\alpha _l}$ represents the complex path gain of $l$th path. The path gain at THz band mainly consists of path loss and molecular absorption, which has been computed in \cite{system_02}. In addition, $\phi _l^{\rm{B}}\left( {\upsilon _l^{\rm{B}}} \right)$  denotes the associated azimuth (elevation) angle of arrival (AoA) and  $\phi _l^{\rm{R}}\left( {\upsilon _l^{\rm{R}}} \right)$ denotes the associated azimuth (elevation) angles of departure (AoD) for $l$th  path. Since THz communication requires a large number of array elements to counter the serious propagation attenuation, the uniform planar array (UPA) structure is more appropriate here. Thus, the normalized array response for the ${N_x}{N_y}$-elements of UPA can be expressed as
\begin{align}\label{system09}
\begin{array}{l}
{\bf{a}}\left( {\phi ,\upsilon } \right) = {{\bf{a}}_x}\left( \phi  \right) \otimes {{\bf{a}}_y}\left( \upsilon  \right)
\end{array},
\end{align}
where ${{\bf{a}}_x}\left( \phi  \right)$  indicates the array response in the azimuth direction and ${{\bf{a}}_y}\left( \upsilon  \right)$  indicates the array response in the elevation direction, which can be concretely described as
\begin{subequations}\label{system10}
\begin{align}
{{\bf{a}}_x}\left( \phi  \right) &= \frac{1}{{\sqrt {{N_x}} }}{\left[ {1, \cdots ,e^{\frac{{j2\pi d{n_x}}}{\lambda }\sin (\phi )}, \cdots } \right]^T},\\
{{\bf{a}}_y}\left( \upsilon  \right) &= \frac{1}{{\sqrt {{N_y}} }}{\left[ {1, \cdots ,e^{\frac{{j2\pi d{n_y}}}{\lambda }\sin (\upsilon )}, \cdots } \right]^T}, \label{system10}
\end{align}
\end{subequations}
where ${n_x} \in \left[ {1,{N_x}} \right]$, ${n_y} \in \left[ {1,{N_y}} \right]$, $\lambda$ is the signal wavelength, and $d$ represents the spacing of adjacent THz array antennas.

Similar to the BS-RIS channel, the RIS-user channel of $k$th  user can be written as
\begin{align}\label{system12}
{{\bf{H}}_k} = \sqrt {\frac{{{N_t}M}}{{{L_{\rm{U}}} + 1}}} \sum\limits_{l = 0}^{{L_{\rm{U}}}} {{\beta _l}{{\bf{a}}_{\rm{R}}}\left( {\omega _l^{\rm{R}},\varsigma _l^{\rm{R}}} \right){\bf{a}}_{\rm{U}}^H\left( {\omega _l^{\rm{U}},\varsigma _l^{\rm{U}}} \right)},
\end{align}
where $k \in \left[ {1,K} \right]$. It is worth noting that the acquisition of CSI is demanded before executing the beam selection process in this paper\cite{system_02a}.

\subsection{Problem Formulation}
With regard to the system model and channel model mentioned above, the sum-rate of the RIS-aided THz MU-MIMO system can be formulated as
\begin{align}\label{system13}
R = \sum\limits_{k = 1}^K {{{\log }_2}\left( {1 + {\gamma _k}} \right)},
\end{align}
where ${\gamma _k}$ is the signal-to-interference-plus-noise ratio (SINR) of $k$th user. According to (\ref{system07}), ${\gamma _k}$ can be written as
\begin{align}\label{system14}
{\gamma _k} &= \frac{{{{\left( {{\bf{s}}{{\bf{s}}^H}} \right)}_{k,k}}}}{{{{\left[ {{{\left( {{{{\bf{\bar {H}}}}^H}{\bf{\bar {H}}}} \right)}^{ - 1}}{{{\bf{\bar {H}}}}^H}{{\bf{W}}_{\rm{R}}}{\bf{n}}{{\left( {{{\left( {{{{\bf{\bar {H}}}}^H}{\bf{\bar {H}}}} \right)}^{ - 1}}{{{\bf{\bar {H}}}}^H}{{\bf{W}}_{\rm{R}}}{\bf{n}}} \right)}^H}} \right]}_{k,k}}}} \notag \\
&= \frac{{{{\bf{s}}_k}{\bf{s}}_k^H}}{{{{\left[ {{{\left( {{{{\bf{\bar {H}}}}^H}{\bf{\bar {H}}}} \right)}^{ - 1}}{{{\bf{\bar {H}}}}^H}{{\bf{W}}_{\rm{R}}}{\bf{n}}{{\bf{n}}^H}{\bf{W}}_{\rm{R}}^H{\bf{\bar {H}}}{{\left( {{{{\bf{\bar {H}}}}^H}{\bf{\bar {H}}}} \right)}^{ - 1}}} \right]}_{k,k}}}} \notag \\
&= \frac{P}{{{N_t}{N_{\rm{B}}}{N_0}{{\left[ {{{\left( {{{{\bf{\bar {H}}}}^H}{\bf{\bar {H}}}} \right)}^{ - 1}}} \right]}_{k,k}}}},
\end{align}
where ${\left[  \cdot  \right]_{k,k}}$  represents the $k$th column and $k$th row element of a matrix.

By observing (\ref{system14}), the SINR of $k$th user closely depends on the equivalent channel ${\bf{\bar { H}}}$  in (\ref{system04}). Furthermore,the definition of ${\bf{\bar { H}}}$  is involved with the analog beamforming matrices, e.g., ${{\bf{W}}_{\rm{R}}}$, ${\bf{\Phi }}$  and  ${\bf{F}}$. Thus, the main target is to select the optimal analog beam combination from the predefined codebooks to maximize the uplink sum-rate metric. Nevertheless, the matrix inversion operation in (\ref{system14}) results in high complexity. In light of this, we replace the inverse operation by using matrix properties in the following.

\emph{Proposition 1}: Given the non-singular matrix ${{\bf{\bar {H}}}^H}{\bf{\bar {H}}}$ , we can get a low-complexity form as
\begin{align}\label{IAS01}
{\left[ {{{\left( {{{{\bf{\bar {H}}}}^H}{\bf{\bar {H}}}} \right)}^{ - 1}}} \right]_{i,i}} = \frac{{\det \left( {{\bf{\bar {H}}}_{\left\langle {i, \sim } \right\rangle }^H{{{\bf{\bar {H}}}}_{\left\langle { \sim ,i} \right\rangle }}} \right)}}{{\det \left( {{{{\bf{\bar {H}}}}^H}{\bf{\bar {H}}}} \right)}},
\end{align}
where ${{{{\bf{\bar {H}}}}_{\left\langle { \sim ,i} \right\rangle }}}$ (or ${{\bf{\bar { H}}}_{\left\langle {i, \sim } \right\rangle }}$) indicates a submatrix formed by deleting $i$th column (or row) of ${{\bf{\bar {H}}}}$. The proof for (\ref{IAS01}) is provided in Appendix A.

In view of (\ref{system14}) and (\ref{IAS01}), the sum-rate metric in (\ref{system13}) can be further rewritten as
\begin{equation}\label{IAS02}
\begin{aligned}
R = \sum\limits_{k = 1}^K {{{\log }_2}\left( {1 + \frac{{P\det \left( {{\bf{\bar {H}}}_{\left\langle {k, \sim } \right\rangle }^H{{{\bf{\bar {H}}}}_{\left\langle { \sim ,k} \right\rangle }}} \right)}}{{{N_t}{N_{\rm{B}}}{N_0}\det \left( {{{{\bf{\bar {H}}}}^H}{\bf{\bar {H}}}} \right)}}} \right)}.
\end{aligned}
\end{equation}

Hence, the sum-rate optimization problem for hybrid beamforming can be formulated as
\begin{equation}\label{IAS03}
\begin{aligned}
\left\{ {{{\bf{F}}^{\star}},{{\bf{\Phi }}^{\star}},{\bf{W}}_{\rm{R}}^{\star}} \right\}  = & \arg \max \; R \\
{\rm s.t.}\;& {{{\bf{f}}_k} \in {\cal F}}, \;\forall k = 1,2, \cdots ,K, \\
&  {{\bf{\Lambda }}_m} \in {\cal S}, \; \forall m = 1,2, \cdots ,{M_s},\\
&  {{\bf{w}}_n} \in {\cal W}, \; \forall n = 1,2, \cdots ,{N_s},
\end{aligned}
\end{equation}
where ${{\bf{F}}^ {\star} } = [{\bf{f}}_1^ {\star} , \cdots ,{\bf{f}}_K^ {\star}]$, ${{\bf{W}_{\rm{R}}}^ {\star} } = {\rm {diag}}({\bf{w}}_1^ {\star} , \cdots ,{\bf{w}}_{N_s}^ {\star})$ and ${{\bf{\Phi}}^ {\star} } = {\rm {diag}}({\rm {vec}}[{\bf{\Lambda}}_1^ {\star} ,{\bf{\Lambda}}_2^ {\star} , \cdots ,{\bf{\Lambda}}_{M_s}^ {\star}])$. Moreover, ${\cal F}$, ${\cal S}$ and ${\cal W}$ are the predefined analog beam codebooks, all of which are composed of the array response vectors and can be easily generated according to \cite{system_03}. To maximize the sum-rate, we need to find the optimal combination of analog beams at the BS, the RIS and all users concurrently.

The exhaustive search (ES) algorithm can achieve the best sum-rate performance by visiting all the possible analog beams in the feasible sets, but endures the exponential increase of complexity, especially for a large number of subarrays or beam candidates. To avoid the excessive beam search, a low complexity iterative alternating search (IAS) algorithm is employed to select partially better beam combinations. Specifically, the number of beam combinations of the IAS algorithm decreases to $({\left| {\cal F} \right|^K} + {\left| {\cal S} \right|^{{M_s}}} + {\left| {\cal W} \right|^{{N_s}}})$ compared with the ES algorithm that possesses   $({\left| {\cal F} \right|^K} \cdot {\left| {\cal S} \right|^{{M_s}}} \cdot {\left| {\cal W} \right|^{{N_s}}})$ possible beam candidates. It is worth noting that the IAS algorithm still yields tremendous beam selection overhead with the increasing number of subarrays and codebook size. Motivated by this, it is imperative to develop a low complexity beam selection framework that combines with MTL technique.

\section{Multi-Task Learning Based Beam Selection}
In this section, the MTL-ABS framework is investigated to solve problem (\ref{IAS03}) and select the optimal beam combination. According to the dataset that is generated by the IAS algorithm in an offline manner, the proposed MTL-ABS can provide non-convex and non-linear mapping between channel inputs and beam labels. In the prediction stage, the MTL-ABS network is able to predict the optimal beam labels of multiple tasks according to the incoming CSI samples.

\subsection{Overall Architecture of MTL-ABS}
\begin{figure*}[t]
\centering
\includegraphics[width=16 cm]{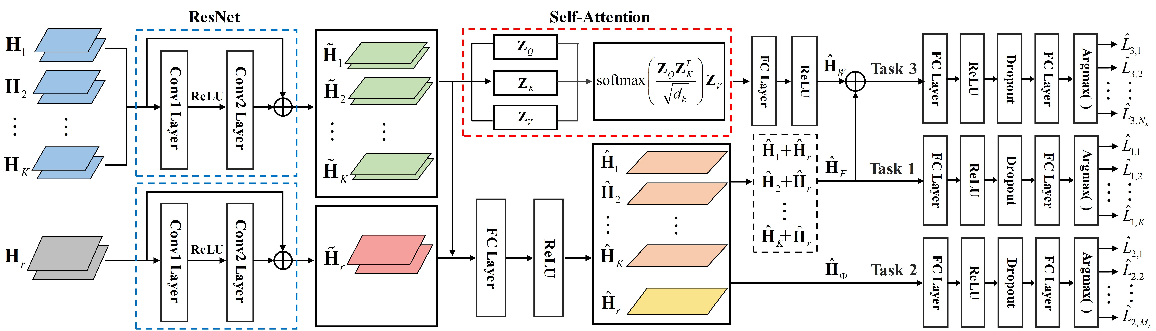}
\caption{Proposed MTL-ABS architecture.}\label{fig2}
\end{figure*}

The overall architecture of the proposed MTL-ABS framework can be illustrated in Fig.\;\ref{fig2}, including shared part and task-specific part. To better understand the working principle of this MTL-ABS framework, we need to summarize some main steps in brief. Firstly, the appropriate MTL-ABS model is established to determine the desired inputs and outputs. Secondly, we use the IAS algorithm to generate the dataset that contains CSI and beam codeword labels. Then, with the sufficient data samples, the MTL-ABS model can be trained to find the mapping relationship between inputs of channel matrices and outputs of beam labels. In addition, given the trained MTL-ABS model, the accurate prediction can be achieved with low complexity via inputting new channel samples. Finally, the prediction accuracy can be utilized to evaluate the training performance of the network model.

\textbf{\emph{1) Shared Representation Design of MTL-ABS}}

The shared part of the MTL-ABS model extracts the implicit features from input channels. In order to fully take advantage of the convolution layers, the 2D channel samples are converted into 3D channel samples by splitting the complex channel matrices into real part $\Re \{ {{\bf{H}}_i}\}$ and imaginary part $\Im \{ {{\bf{H}}_i}\}$, where $i \in {\cal I} = \left\{ {1,2, \cdots ,K} \right\} \cup \left\{ r \right\}$. The input 3D channels are processed by two ResNet blocks and each block contains two convolutional layers: Conv1 and Conv2. Specifically, Conv1 aims to convert the low dimensional channels into high dimensional channels. The output data of Conv1 is performed by the ReLU activation function $\delta \left( x \right) = \max \left( {0,x} \right)$, which enhances the nonlinear relations of network parameters and alleviates the overfitting issue. Then, the function of Conv2 is to restore the high dimensional channels processed by Conv1 to originally low dimensional channels. Although the data dimension between input and output of ResNet remains unchanged, some significant channel features have been captured. Moreover, the ResNet is conducive to solving the network degradation and gradient dissipation problem by adding the identity shortcuts that are parameter-free. Thus, the ResNet in the MTL-ABS model can be expressed as
\begin{align}\label{MTL01}
{{\bf{\tilde H}}_i} = {f_{{\rm{C2}}}}\left( {\delta \left( {{f_{{\rm{C1}}}}\left( {\begin{array}{*{20}{c}}
{{\mathop{\Re}\nolimits} \left\{ {{{\bf{H}}_i}} \right\}}\\
{{\mathop{\Im}\nolimits} \left\{ {{{\bf{H}}_i}} \right\}}
\end{array}} \right)} \right)} \right) + \left( {\begin{array}{*{20}{c}}
{{\mathop{\Re}\nolimits} \left\{ {{{\bf{H}}_i}} \right\}}\\
{{\mathop{\Im}\nolimits} \left\{ {{{\bf{H}}_i}} \right\}}
\end{array}} \right),
\end{align}
where $i \in {\cal I}$ and ${f_{{\rm{C}}}}\left(  \cdot  \right)$  is defined as the convolutional projection function facilitating the residual mapping.

\textbf{\emph{2) Task-Specific Design of MTL-ABS}}

Except for the shared network part in Fig.\;\ref{fig2}, the task-specific network part is also designed for different beam classification tasks by considering special channel characteristics of the communication scenario.

$\bullet$ \emph{Task 1 for} ${{\bf{F}}^{\star}}$: This task is to accomplish the optimal beam selection for all $K$ users. Since the LoS path is blocked by obstacles, we need to direct the beams of all $K$ users towards the RIS.
According to such a practical scenario, it is obvious that the beam selection of $k$th user is strong correlation with its own channel ${{\bf{\hat H}}_k}$ and reflected channel ${{\bf{\hat H}}_r}$. In other words, the analog training of $K$ users is independent of each other, and thus each user can focus on learning its own channel features. Motivated by this important fact, the channel inputs of all $K$ users for task 1 can be written as
\begin{align}\label{MTL02}
{{{\bf{\hat H}}}_F} = {\left[ {{{\left( {{{{\bf{\hat H}}}_r} + {{{\bf{\hat H}}}_1}} \right)}^T}, \cdots ,{{\left( {{{{\bf{\hat H}}}_r} + {{{\bf{\hat H}}}_K}} \right)}^T}} \right]^T},
\end{align}
where  ${{\bf{\hat H}}_F}$ can be fed to the FC layers to get the desired beam labels $\left\{ {{L_{1,k}}} \right\}_{k = 1}^K$. In addition, the dropout layer is deployed between two FC layers, which can effectively alleviate the occurrence of overfitting and achieve the regularization effect to a certain extent.

$\bullet$ \emph{Task 2 for }${{\bf{\Phi }}^{\star}}$: Task 2 is to realize the optimal beam selection for multiple subarrays at the RIS. Generally, RIS is installed in the middle of the BS and users to build the virtual LoS path,  so the passive beamforming design at the RIS needs to learn all the related channels, which is greatly different from task 1. In this regard, the input channels for task 2 can be given by  ${{{\bf{\hat H}}}_\Phi } = {[{{{\bf{\hat H}}}_r},{{{\bf{\hat H}}}_1},{{{\bf{\hat H}}}_2}, \cdots ,{{{\bf{\hat H}}}_K}]^T}$, and we directly feed the channel matrix into the FC layers. Then, the desired labels $\left\{ {{L_{2,m}}} \right\}_{m = 1}^{{M_s}}$ for the RIS can be obtained accordingly.

$\bullet$ \emph{Task 3 for }${\bf{W}}_{\rm{R}}^{\star}$: Task 3 aims to tackle the analog beam selection problem at the BS. Nevertheless, compared with aforementioned tasks, task 3 is hard to find the implicative relationship between input channels and beam labels. In light of this challenge, we are unable to directly design the network structure for task 3 by simply applying the channel allocation strategy. To this end, the self-attention mechanism, which does well in capturing the internal correlation of data features, is used to adaptively reallocate the channel weights and couples with the beam selection problem at the BS. Given the input channel matrix, we need to calculate the random matrices of query ${{\bf{Z}}_Q}$, key ${{\bf{Z}}_K}$  and value ${{\bf{Z}}_V}$. Since each random matrix is updated during the training stage, we user a single FC layer to replace a specific random matrix. Hence, ${{\bf{Z}}_Q}$, ${{\bf{Z}}_K}$  and ${{\bf{Z}}_V}$ can be respectively expressed as
\begin{subequations}\label{MTL03}
\begin{align}
{{\bf{Z}}_Q} &= {f_{{\rm{FC\_Q}}}}\left( {{{{\bf{\tilde H}}}_i}} \right),i \in {\cal I},\\
{{\bf{Z}}_K} &= {f_{{\rm{FC\_K}}}}\left( {{{{\bf{\tilde H}}}_i}} \right),i \in {\cal I},\\
{{\bf{Z}}_V} &= {f_{{\rm{FC\_V}}}}\left( {{{{\bf{\tilde H}}}_i}} \right),i \in {\cal I}, \label{MTL03}
\end{align}
\end{subequations}
where ${f_{{\rm{FC}}}}\left(  \cdot  \right)$ denotes the linear mapping operation of a single FC layer. After being processed by the FC layer and ReLU, the output matrix of the self-attention can be given by
\begin{align}\label{MTL04}
{{\bf{\hat H}}_W} = \delta \left( {{f_{{\rm{FC}}}}\left( {{\rm{softmax}}\left( {\frac{{{{\bf{Z}}_Q}{\bf{Z}}_K^T}}{{\sqrt {{d_K}} }}} \right){{\bf{Z}}_V}} \right)} \right),
\end{align}
where $d_K$  indicates the dimension of  ${{\bf{Z}}_K}$ (e.g., $d_K=1024$) and  ${\rm{softmax}}\left(  \cdot  \right)$ is defined as a typical soft-max function. Considering the effect of the BS-RIS channel ${{\bf{\hat H}}_r}$, we treat the compound term $({{\bf{\hat H}}_W} + {{\bf{\hat H}}_F})$ as the input of task 3, and then the optimal beam candidate $\left\{ {{L_{3,n}}} \right\}_{n = 1}^{{N_s}}$  can be learned at the BS.

\subsection{Dataset Generation}
As mentioned before, the beam selection dataset is generated by the IAS algorithm, which is composed of channel inputs and beam label outputs. In detail, the inputs of the MTL-ABS network contain ${\bf{H}}_r^{(t)}$  and $\{ {\bf{H}}_k^{(t)}\} _{k = 1}^K$, and the output labels can be denoted as  $\{ L_{1,k}^{(t)}\} _{k = 1}^K$, $\{ L_{2,m}^{(t)}\} _{m = 1}^{{M_s}}$  and  $\{ L_{3,n}^{(t)}\} _{n = 1}^{{N_s}}$, where $t$ denotes the index of the data samples. These desired labels correspond to the optimal analog beams  $\{ {\bf{f}}_k^{(t)}\} _{k = 1}^K$,  $\{ {\bf{\Lambda }}_m^{(t)}\} _{m = 1}^{{M_s}}$ and  $\{ {\bf{w}}_n^{(t)}\} _{n = 1}^{{N_s}}$, respectively. By aggregating the input channel matrices and the output labels together, a complete dataset can be generated to train the MTL-ABS model. Furthermore, we divide the whole dataset into training dataset ${\cal D}_{\text {tra}}$  and validation dataset ${\cal D}_{\text {val}}$. In this paper, there are total $6 \times {10^4}$  data samples, including $5 \times {10^4}$  training samples and  ${10^4}$ validation samples. Then, we transform the original data samples into the mini-batch form that is convenient to implement the multi-threads parallel processing. Thus, the training time can be greatly curtailed and the appropriate batch size also accelerates network convergence.

It is worth noting that the dataset possesses some distinctive characteristics. On the one hand, since the sum-rate metric of the RIS enabled THz MU-MIMO system takes the noise variance into consideration, the dataset generated by the IAS algorithm has the ability to combat the imperfection of the new input channels caused by AWGN. On the other hand, the channel samples are randomly generated in a given communication environment, and thus the unexpected behavior of the wireless propagation channels can be handled as long as the number of data samples is large enough. In view of these advantages brought by the dataset, the proposed MTL-ABS scheme is able to acquire more robust sum-rate performance.

\subsection{Training Process}
In order to improve the training efficiency, the training dataset are packaged as the form of batches and each batch is comprised of ${N_{bat}} = 16$  data samples. In addition, the adaptive moment estimation (Adam) is considered as the optimizer of our proposed MTL-ABS model, which is an efficient stochastic optimization method that only considers first-order gradients with little memory requirement \cite{system_05}. Specifically, the mean and square coefficients for computing gradients in the Adma optimizer is set as (0.8, 0.999). For the sake of better training performance, the MTL-ABS model chooses different learning rate values as the initial states. By using the scheduler learning strategy, the learning rate $\eta$  usually becomes smaller along with the increasing number of epochs ${N_{{\rm{epo}}}}$. Moreover, the loss function is another indispensable criterion to measure the fitness of parameter update for the network model. Since the multiple analog beam selection problem belongs to the discrete classification problem, the cross-entropy loss function is an appropriate choice to guarantee high classification accuracy of the training process. For the training dataset  ${\cal D}_{\text {tra}}$, the overall loss function is composed of the loss summation of multiple learning tasks, which can be written as
\begin{align}\label{MTL05}
{{\cal L}_{{\rm{MTL}}}} = \frac{1}{U}\sum\limits_{u = 1}^U {{{\cal L}_u}\left( {{\bf{G}},{{\bf{Q}}_u}} \right)},
\end{align}
where ${{\cal L}_u}$ indicates the loss function of $u$th task and the task index satisfies $u \in {\cal U} = \{ 1,2, \cdots ,U\}$. In addition, ${\bf{G}}$ and ${{\bf{Q}}_u}$ are the shared and task-specific network weights. The main goal is to minimize the overall training loss by optimizing the network parameters, which can be given by
\begin{align}\label{MTL051}
\left\{ {{\bf{G}}^{\star},{{\bf{Q}}_{u \in {\cal U}}^{\star}}} \right\} = \mathop {\arg \min }\limits_{{\bf{G}},{{\bf{Q}}_{u \in {\cal U}}}} {{\cal L}_{{\rm{MTL}}}}\left( {{\bf{G}},{{\bf{Q}}_{u \in {\cal U}}}} \right).
\end{align}

Once the problem (\ref{MTL051}) is solved, the optimized network weights can be obtained. Then, the cross-entropy loss for $u$th learning task can be specifically calculated as
\begin{align}\label{MTL06}
{{\cal L}_u} =  - \frac{1}{{\left| {{{\cal D}_{{\rm{tra}}}}} \right|{p_u}}}\sum\limits_{t = 1}^{\left| {{{\cal D}_{{\rm{tra}}}}} \right|} {\sum\limits_{j = 1}^{{p_u}} {L_{u,j}^{(t)}\log \left( {\frac{{{e^{{\bf r}_{u,j}^{(t)}\left( {L_{u,j}^{(t)}} \right)}}}}{{\sum\nolimits_{i = 1}^{\left| {{{\cal C}_u}} \right|} {{e^{{\bf r}_{u,j}^{(t)}\left( i \right)}}} }}} \right)} },
\end{align}
where $p_u$ is the number of outputs for $u$th task and each output contains $|{{\cal C}_u}|$ categories, e.g., ${|{\cal F}|}$, ${|{\cal S}|}$ and ${|{\cal W}|}$. Moreover,  ${\bf{r}}_{u,j}^{(t)} = [{\bf{r}}_{u,j}^{(t)}(1), \cdots ,{\bf{r}}_{u,j}^{(t)}(|{{\cal C}_u}|)]$ represents the vector of $j$th output for $u$th task that contains raw and unnormalized scores, and we can further get $\hat L_{u,j}^{(t)} = \arg \max ({\bf{r}}_{u,j}^{(t)})$ accordingly.

\subsection{Validation Process}
After training the MTL-ABS model, we import the validation dataset ${\cal D}_{\text {val}}$  into the trained model to test the network generalization capability. To avoid the similarity between ${\cal D}_{\text {val}}$ and ${\cal D}_{\text {tra}}$, the channel parameters (e.g., path gains, AoAs and AoDs) are randomly generated for the validation samples. The new channel samples are fed into the trained model in the case of  ${N_{bat}} = 1$, and then the prediction accuracy can be obtained by comparing the prediction results with the original labels generated by the IAS algorithm. The calculation formula of the overall prediction accuracy rate can be given by
\begin{align}\label{MTL09}
{{\cal A}_{{\rm{MTL}}}} = \frac{1}{U}\sum\limits_{u = 1}^U {{{\cal A}_u}\left( {{\bf{G}}^{\star},{{\bf{Q}}_{u}^{\star}}} \right)},
\end{align}
where ${{\cal A}_{u}}$ denotes the prediction accuracy values for $u$th task, which can be further formulated as
\begin{align}\label{MTL10}
{{\cal A}_u} = \frac{1}{{\left| {{{\cal D}_{{\rm{val}}}}} \right|{p_u}}}\sum\limits_{t = 1}^{\left| {{{\cal D}_{{\rm{val}}}}} \right|} {\sum\limits_{j = 1}^{{p_u}} {{\rm{equal}}\left( {L_{u,j}^{(t)},\hat L_{u,j}^{(t)}} \right)} },
\end{align}
where ${\rm{equal}}\left(  \cdot  \right)$  indicates the comparison function that the function value is set to 1 if the two input elements are equal otherwise the function value is set to 0. Depending on the accuracy criterion, the appropriate parameters of the MTL-ABS model can be validated as well as the training effectiveness.

\section{Convergence and Complexity Analysis of MTL-ABS Model}
In this section, we explore the blockwise convergence our proposed MTL-ABS framework and provide the complexity analysis of considered beam selection algorithms.

\subsection{Blockwise Convergence Analysis for MTL-ABS}
\begin{figure}[!t]
\centering
\includegraphics[width=8cm]{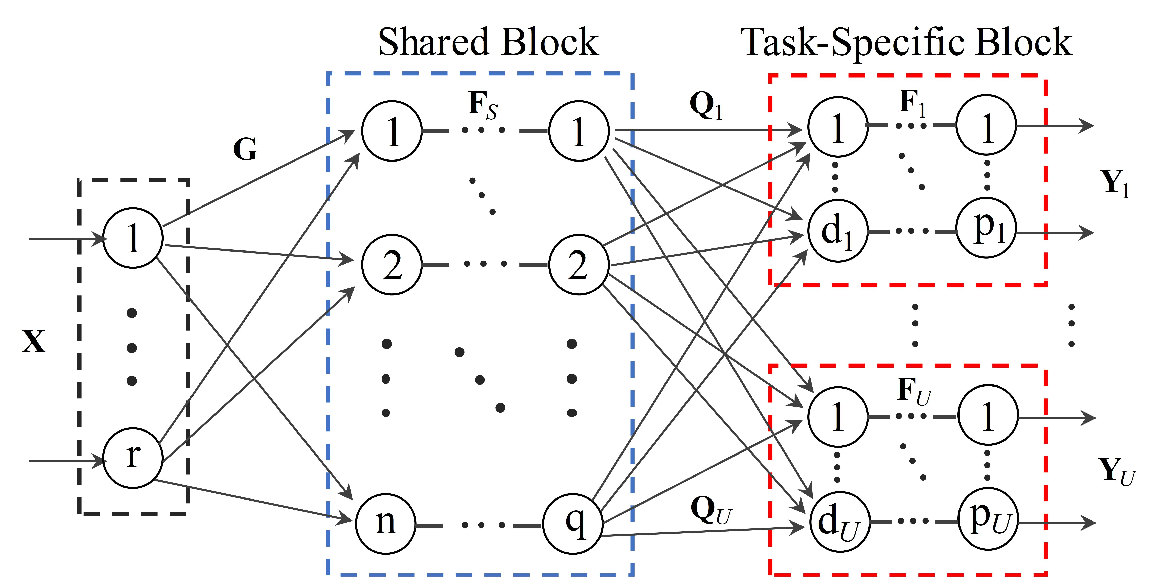}
\caption{Blockwise structure of proposed MTL-ABS.}\label{figBC}
\end{figure}

To the best of our knowledge, it is difficult to theoretically analyze a complex MTL network. In this regard, we originally develop a blockwise approach to analyze the convergence of our proposed MTL-ABS network, which is divided into different function blocks, e.g., shared block and task-specific blocks, as shown in Fig. \ref{figBC}. Each function block represents a distinctive sub-network, and these function blocks can be combined in a required manner to realize different classification tasks. More importantly, we treat each function block as a black box during the theoretical analysis process. It is worth noting that each black box has the ability to approximate any desired continuous function or classify any disjoint regions. The internal structure of each black box can be complicatedly designed based on the required function, but the connections among these sub-network blocks remain the linear mapping property.

Suppose that we have $U$ tasks and the training dataset can be formulated as  ${\cal D}_{\text {tra}} = \{ {\bf{X}};{{\bf{Y}}_1},{{\bf{Y}}_2}, \cdots ,{{\bf{Y}}_U}\}$, where ${\bf{X}} = {[{{\bf{x}}_1},{{\bf{x}}_2}, \cdots ,{{\bf{x}}_V}]^T} \in {{\mathbb C}^{V \times r}}$  is the common inputs of $V$ samples and ${{\bf{Y}}_u} = {[{{\bf{y}}_{u,1}},{{\bf{y}}_{u,2}}, \cdots ,{{\bf{y}}_{u,V}}]^T} \in {{\mathbb C}^{V \times {p_u}}}$  denotes the outputs of $V$ samples for $u$th task. Hence, the optimization problem of the blockwise MTL-ABS architecture can be formulated as
\begin{align}\label{BWC01}
\left\{ {{{\bf{G}}^ {\star} },{{\bf{Q}}^ {\star} }} \right\}: = \mathop {{\arg \min } }\limits_{{\bf{G}},{\bf{Q}}} {f_0}\left( {{\bf{G}},{\bf{Q}}} \right) + {f_S}\left( {\bf{G}} \right) + {f_T}\left( {\bf{Q}} \right),
\end{align}
where
\begin{equation}
\begin{aligned}
{f_0}\left( {{\bf{G}},{\bf{Q}}} \right) = \sum\limits_{u = 1}^U {\frac{1}{2}\left\| {{\bf{XG}}{{\bf{F}}_S}{{\bf{Q}}_u}{{\bf{F}}_u} - {{\bf{Y}}_u}} \right\|_F^2}, \\
{f_S}\left( {\bf{G}} \right) = \frac{{{\rho _1}}}{2}\left\| {\bf{G}} \right\|_F^2, \;\;\; {f_T}\left( {\bf{Q}} \right) = \frac{{{\rho _2}}}{2}\left\| {\bf{Q}} \right\|_F^2,
\end{aligned}
\end{equation}
and ${\bf{G}} \in {{\mathbb C}^{r \times n}}$ is the weights of shared network part as shown in Fig. \ref{figBC}. ${{\bf{Q}}_u} \in {{\mathbb C}^{q \times {d_u}}}$ indicates the network weights of $u$th task and we further define ${\bf{Q}} = {[{\bf{Q}}_1^T,{\bf{Q}}_2^T, \cdots ,{\bf{Q}}_U^T]^T}$. ${{\bf{F}}_S} \to {{\mathbb R}^{n \times q}}$  and ${{\bf{F}}_u} \to {{\mathbb R}^{{d_u} \times {p_u}}}(u = 1,2, \cdots ,U)$ represent the inner mapping functions of the shared block and the $u$th  task-specific block, and these desired mapping functions can be obtained by designing the required network architecture based on the practical application (e.g., MTL-ABS). In addition, the parameters ${\rho _1} > 0$  and ${\rho _2} > 0$ represent the penalty coefficients. The penalty terms ${{{\rho _1}||{\bf{G}}||_F^2} \mathord{\left/ {\vphantom {{{\rho _1}||{\bf{G}}||_F^2} 2}} \right. \kern-\nulldelimiterspace} 2}$  and ${{{\rho _2}||{\bf{Q}}||_F^2} \mathord{\left/ {\vphantom {{{\rho _2}||{\bf{Q}}||_F^2} 2}} \right. \kern-\nulldelimiterspace} 2}$  stem from the Bartlett's theory \cite{BWC_01} that smaller norms of the weights tend to own better network generalization performance.

The main goal of problem (\ref{BWC01}) is to find the optimal ${\bf{G}}$  and  ${\bf{Q}}$ by minimizing the training loss. It can be observed that problem (\ref{BWC01}) is bi-convex with ${\bf{G}}$  and  ${\bf{Q}}$, and thus the alternating optimization (AO) can be employed. Fixed ${\bf{Q}}^i$, the objective (\ref{BWC01}) for optimizing ${{\bf{G}}}$  can be reformulated as
\begin{align}\label{BWC02}
{{\bf{G}}^{i + 1}}: = \mathop {\arg \min }\limits_{\bf{G}} {f_0}\left( {{\bf{G}},{{\bf{Q}}^{i + 1}}} \right) + {f_S}\left( {\bf{G}} \right),
\end{align}
where superscript $i$ denotes the $i$th iteration of AO method. It can be observed that the problem (\ref{BWC02}) is convex in terms of ${\bf{G}}$. Generally, by taking the derivative as zero,  ${{\bf{G}}^{i + 1}}$ can be written as
\begin{align}\label{BWC03}
& {\rm {vec}}\left( {{{\bf{G}}^{i + 1}}} \right) \notag \\
&= {\left( {\sum\limits_{u = 1}^U {\left( {{{\bf{F}}_S}{\bf{Q}}_u^i{{\bf{F}}_u}{\bf{F}}_u^T{\bf{Q}}_u^{i,T}{\bf{F}}_S^T} \right) \otimes \left( {{{\bf{X}}^T}{\bf{X}}} \right)}  + {\rho _1}{{\bf{I}}_{nr}}} \right)^{ - 1}}\notag \\
&\times {\rm {vec}}\left( {\sum\limits_{u = 1}^U {{{\bf{X}}^T}{{\bf{Y}}_u}{\bf{F}}_u^T{\bf{Q}}_u^{i,T}} {\bf{F}}_S^T} \right).
\end{align}

With fixed  ${{\bf{G}}^{i + 1}}$, $U$ tasks in problem (\ref{BWC01}) can be solved separately. Thus, the $u$th task can be formulated as
\begin{align}\label{BWC04}
{{\bf{Q}}^{i + 1}_u}: = \mathop {\arg \min }\limits_{{\bf{Q}}_u} {f_0}\left( {{{\bf{G}}^{i + 1}},{\bf{Q}}}_u \right) + {f_T}\left( {\bf{Q}}_u \right).
\end{align}

Similar to problem (\ref{BWC02}), problem (\ref{BWC04}) is also convex with regard to ${{\bf{Q}}_u}$. Hence, ${\bf{Q}}_u^{i + 1}$ can be given by
\begin{align}\label{BWC05}
& {\rm {vec}}\left( {{\bf{Q}}_u^{i + 1}} \right) \notag \\
&= {\left( {\left( {{{\bf{F}}_u}{\bf{F}}_u^T} \right) \otimes \left( {{\bf{F}}_S^T{{\bf{G}}^{i + 1,T}}{{\bf{X}}^T}{\bf{X}}{{\bf{G}}^{i + 1}}{{\bf{F}}_S}} \right) + {\rho _2}{{\bf{I}}_{q{d_u}}}} \right)^{ - 1}} \notag \\
& \times {\rm {vec}}\left( {{\bf{F}}_S^T{{\bf{G}}^{i + 1,T}}{{\bf{X}}^T}{{\bf{Y}}_u}{\bf{F}}_u^T} \right).
\end{align}

By iteratively proceeding (\ref{BWC03}) and (\ref{BWC05}), ${\bf{G}}$  and ${\bf{Q}}$ can be sequentially optimized. However, the major challenge faced by the AO method is to prove the convergence, which is provided as follow.

\emph{Proposition 2}: With the iteration number reaching the condition $i \to \infty$ , the network weight sequence  $\{ {{\bf{G}}^i},{{\bf{Q}}^i}\}$ of the blockwise MTL-ABS architecture converges to a stationary point $\{ {{\bf{G}}^{ \star}},{{\bf{Q}}^{ \star}}\}$.

The proof for \emph{Proposition 2} is provided in Appendix B.

It should be noted that our proposed MTL-ABS framework are consistent with the blockwise MTL network by dividing the MTL-ABS network into different function blocks. Hence, the proposed MTL-ABS structure can be cast as a special case of the blockwise MTL structure and \emph{Proposition 2} is capable of guaranteeing the convergence of our proposed MTL-ABS framework.

\subsection{Complexity Analysis}
To begin with, the ES algorithm and the IAS algorithm visit $(|{\cal F}{|^K}|{\cal S}{|^{{M_s}}}|{\cal W}{|^{{N_s}}})$ and $( {{{{\left| {\cal F} \right|}^K} + {{\left| {\cal S} \right|}^{{M_s}}} + {{\left| {\cal W} \right|}^{{N_s}}}}})$ possible beam combinations, respectively. For one  candidate, the complexity of computing ${\bf{\bar {H}}}$ can be written as
\begin{align}\label{MTL13}
{{\cal O}_{\rm{1}}}\left( {K{N_s}\left( {{N_r}M + {M^2} + M{N_t} + {N_t}} \right)} \right).
\end{align}

 Given ${\bf{\bar {H}}}$, the complexity of calculating the compound term $\det ({\bf{\bar {H}}}_{\left\langle {k, \sim } \right\rangle }^H{{{\bf{\bar {H}}}}_{\left\langle { \sim ,k} \right\rangle }})/\det ({{{\bf{\bar {H}}}}^H}{\bf{\bar {H}}})$ can be written as
\begin{align}\label{MTL14}
{{\cal O}_2}\left( {{K^2}{N_s} + {{\left( {K - 1} \right)}^2}{N_s} + {K^3} + {{\left( {K - 1} \right)}^3} + 1} \right).
\end{align}

Then, the complexity of calculating the sum-rate metric for one analog beam candidate can be expressed as
\begin{align}\label{MTL15}
{{\cal O}_3}\left( \begin{array}{l}
{K^2}{N_s}M\left( {{N_r} + M} \right)\\
 + {K^2}{N_s}{N_t}\left( {M + 1} \right)\\
 + 6K + {K^3}\left( {{N_s} + K} \right)\\
 + K{\left( {K - 1} \right)^2}\left( {{N_s} + K + 1} \right)
\end{array} \right).
\end{align}

Hence, the total complexity of the ES algorithm and the IAS algorithm can be respectively given by
\begin{align}\label{MTL16}
& {{\cal O}_{{\rm{ES}}}}\left( {{{\left| {\cal F} \right|}^K}{{\left| {\cal S} \right|}^{{M_s}}}{{\left| {\cal W} \right|}^{{N_s}}} \cdot {{\cal O}_3}} \right), \\
& {{\cal O}_{{\rm{IAS}}}}\left( {{t_{\max }}\left( {{{\left| {\cal F} \right|}^K} + {{\left| {\cal S} \right|}^{{M_s}}} + {{\left| {\cal W} \right|}^{{N_s}}}} \right) \cdot {{\cal O}_3}} \right),
\end{align}
where ${t_{\max }}$  indicates the number of maximum iterations.

Next, the complexity of our proposed MTL-ABS framework is analyzed in detail. Without loss of generality, the complexity of offline training can be neglected, and thus we focus on the computation overhead of online prediction. Once the network model is trained, the major complexity stems from the multiplications between the input parameters and the layer weights. For two ResNet blocks that separately process the input channel information, the complexity depends on the convolutional layers \cite{system_06}, which can be written as
\begin{align}\label{MTL17}
{{\cal O}_{{\rm{ResNet}}}}\left( \begin{array}{l}
K\sum\limits_{l = 1}^{{L_{C1}}} {{C_{in,l}}{C_{out,l}}C_{ker,l}^2{D_{1,l}}{D_{2,l}}} \\
 + \sum\limits_{l = 1}^{{L_{C2}}} {{{\bar C}_{in,l}}{{\bar C}_{out,l}}\bar C_{ker,l}^2{{\bar D}_{1,l}}{{\bar D}_{2,l}}}
\end{array} \right),
\end{align}
where $l$ is the index of a given convolutional layer, and  ${L_{C1}}$ (or ${L_{C2}}$) is the number of convolutional layers for the first (or second) ResNet. ${C_{ker,l}}$  and ${\bar C_{ker,l}}$  indicate the kernel size in different ResNet blocks.  ${C_{in,l}}$ (or ${\bar C_{in,l}}$) and  ${C_{out,l}}$ (or ${\bar C_{out,l}}$) denote a number of input and output channels for $l$th  convolutional layer in the first (or second) ResNet, respectively. In addition, ${D_{1,l}}$ (or ${\bar D_{1,l}}$ ) and ${D_{2,l}}$  (or ${\bar D_{2,l}}$) are the spatial size of the output feature map. Particularly, in the case of $l=1$, the output feature map can be given by
\begin{align}\label{MTL18}
\begin{array}{l}
{D_{1,1}} = M - {C_{ker,1}} + 3,\;\;{D_{2,1}} = {N_t} - {C_{ker,2}} + 3,\\
{{\bar D}_{1,1}} = {N_r} - {{\bar C}_{ker,1}} + 3,\;\;{{\bar D}_{2,1}} = M - {{\bar C}_{ker,2}} + 3.
\end{array}
\end{align}

Except for the complexity of the ResNet blocks, the complexity brought by FC layers is another important aspect. Specifically, the complexity of ${l_{{\rm{FC}}}}$th  FC layer involves the product of the outputs of $\left( {{l_{{\rm{FC}}}} - 1} \right)$th  layer and the weights of ${l_{{\rm{FC}}}}$th  layer, where ${l_{{\rm{FC}}}} = 1,2, \cdots ,{L_{{\rm{FC}}}}$. Hence, the complexity of ${L_{{\rm{FC}}}}$ FC layers and input-output layers in our trained model can be written as
\begin{align}\label{MTL19}
{{\cal O}_{{\rm{FC}}}}\left( \begin{array}{l}
2KM{N_t}{E_1} + 2M{N_r}{E_1}\\
 + \sum\limits_{{l_{{\rm{FC}}}} = 2}^{{L_{{\rm{FC}}}}} {{E_{{l_{{\rm{FC}}}} - 1}}{E_{{l_{{\rm{FC}}}}}}}  + K{M_s}{N_s}{E_{{L_{{\rm{FC}}}}}}
\end{array} \right),
\end{align}
where ${E_{{l_{{\rm{FC}}}}}}$ represents the number of neural units for ${l_{{\rm{FC}}}}$th layer. Ultimately, the overall prediction complexity of the proposed MTL-ABS scheme can be expressed as
\begin{align}\label{MTL20}
{{\cal O}_{{\text{MTL-ABS}}}}\left( {{{\cal O}_3} + {{\cal O}_{{\rm{FC}}}} + {{\cal O}_{{\rm{ResNet}}}}} \right).
\end{align}

\section{Simulation Results}
In this section, we first investigate the appropriate learning parameters of our proposed MTL-ABS model. To demonstrate the effectiveness of the MTL-ABS framework, we also explore the sum-rate performance comparisons among diverse beam selection schemes as well as complexity comparisons.

\subsection{Simulation Settings}
We consider an uplink RIS-enabled THz MU-MIMO system with different parameter settings for diverse analog beam selection schemes. The working frequency of the RIS is set as $f=1.6$ THz. The reflecting amplitude is set to $0.8$ and the maximum phase response is set as $306.82$ degrees according to \cite{introduction_18}. In addition, we assume that the LoS paths of $K$ users are blocked by the obstacles, and the RIS provides the virtual LoS links among the BS and users. Considering the sparse nature of THz channel,  ${{\bf{H}}_r}$ and ${{\bf{H}}_k}$ contain ${L_{\rm{B}}} = 3$ and ${L_{\rm{U}}} = 3$  propagation paths, including one LoS path and two NLoS paths. To calculate the complex path gain, the molecular absorption coefficient and the reflection coefficient of the obstacle are $\kappa \left( f \right) = 0.2$  and $\xi \left( f \right) = {10^{ - 6}}$  \cite{system_02}. The AoAs and AoDs of the THz channel model are selected from  $\left( { - \pi ,\pi } \right)$ following the uniform distribution. In addition, the communication distance of BS-RIS link is set to ${d_0} = 20\;m$  while the users are randomly deployed in a circular region with the diameter as  ${d_u} = 24\;m$. Under the parameter settings, the simulation results of the heuristic beam search algorithms are performed based on 1000 random channel realizations.

\subsection{Evaluation of MTL-ABS Model}

\begin{figure}[!t]
\centering
\subfloat[]{
\begin{minipage}[t]{1\linewidth}
\centering
\includegraphics[width=6.8cm]{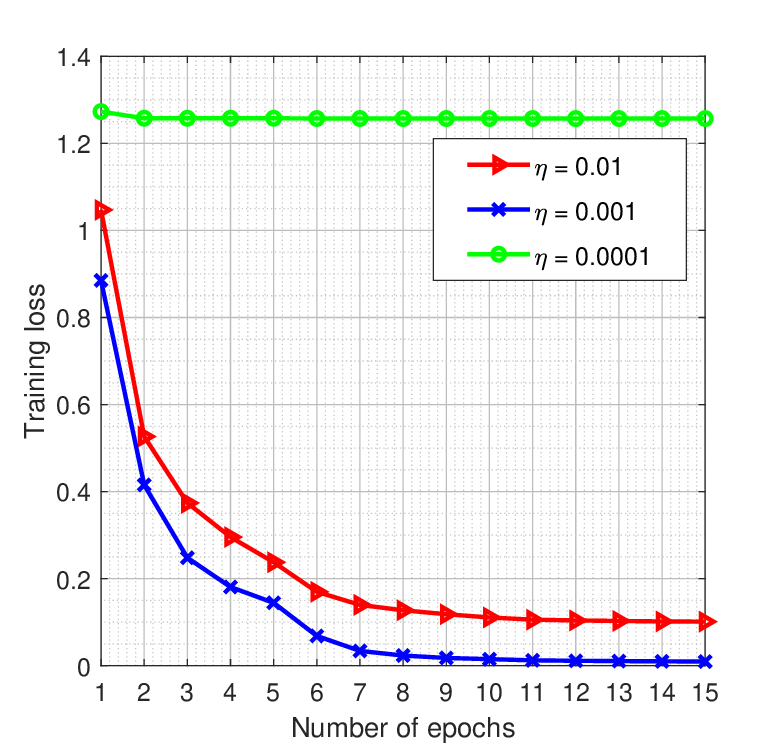}
\end{minipage}%
}\\
\subfloat[]{
\begin{minipage}[t]{1\linewidth}
\centering
\includegraphics[width=6.8cm]{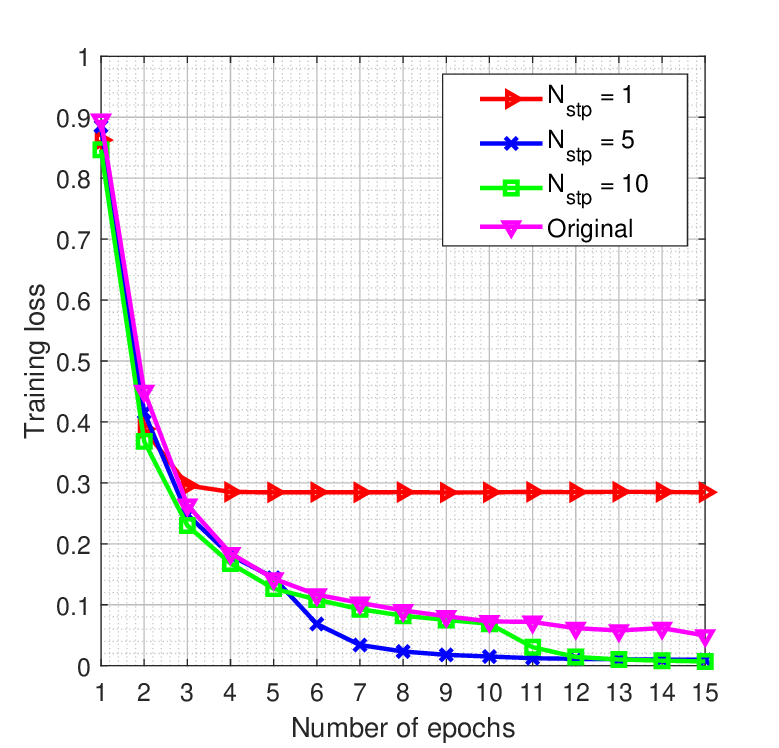}
\end{minipage}
}
\caption{Training loss versus number of epochs: (a) different learning rates; (b) different step sizes.}\label{fig3}
\end{figure}

To begin with, the parameter tuning is one of the most significant steps for a new-built network model, which dominates the final sum-rate performance of the RIS-aided THz MU-MIMO system. Here we consider the system settings as $N_r=128$,  $N_s=4$, $M=128$, $M_s=2$, ${N_t} = 16$, $K=4$. Fig. \,\ref{fig3} depicts the training loss of our proposed model with the increasing number of epochs in terms of different learning rates $\eta $  and step sizes ${N_{{\rm{stp}}}}$. Specifically, as shown in Fig. \,\ref{fig3}(a), the overall trends of the training loss curves are decreasing along with the training epochs, and the training loss is greatly affected by diverse learning rate values. When the learning rate is set to  $\eta=0.001 $, we can acquire the optimal training result for our proposed MTL-ABS model. In addition, from Fig. \,\ref{fig3}(a), we can also observe that the training loss curve with $\eta=0.001$ has basically converged with ${N_{{\rm{epo}}}} = 10$. Given the initial learning rate $\eta=0.001$, we also employ the learning rate scheduler function (e.g., StepLR mechanism) to dynamically adjust the learning rate during the training process. By using the StepLR mechanism, the updated learning rate can be described as ${\eta _{{\rm{new}}}} = \eta  \times {\mu ^q}$, where $\mu$  denotes the multiplication factor for updating the learning rate and $q$ can be defined as $q = {{{N_{{\rm{epo}}}}} \mathord{\left/ {\vphantom {{{N_{{\rm{epo}}}}} {{N_{{\rm{stp}}}}}}} \right. \kern-\nulldelimiterspace} {{N_{{\rm{stp}}}}}}$. In detail, $\mu = 0.1$ is usually configured  while we simulate diverse values of  ${N_{{\rm{stp}}}}$ to find the optimal step size as shown in Fig. \,\ref{fig3}(b). In contrast with original scheme with fixed $\eta$, the training loss curves with ${N_{{\rm{stp}}}} = 5$  and  ${N_{{\rm{stp}}}} = 10$ can be greatly depressed, and the curve with ${N_{{\rm{stp}}}} = 5$ converges faster. Instead, the training loss curve with ${N_{{\rm{stp}}}} = 1$  performs the worst. Due to the StepLR mechanism, there is a sharp sum-rate improvement appearing at  ${N_{{\rm{epo}}}} = 5$. Therefore, the StepLR mechanism with appropriate step size (e.g., ${N_{{\rm{stp}}}} = 5$) contributes to declining the training loss and shortening the training time for our proposed MTL-ABS model.

\begin{figure}[!t]
\centering
\includegraphics[width=6.8cm]{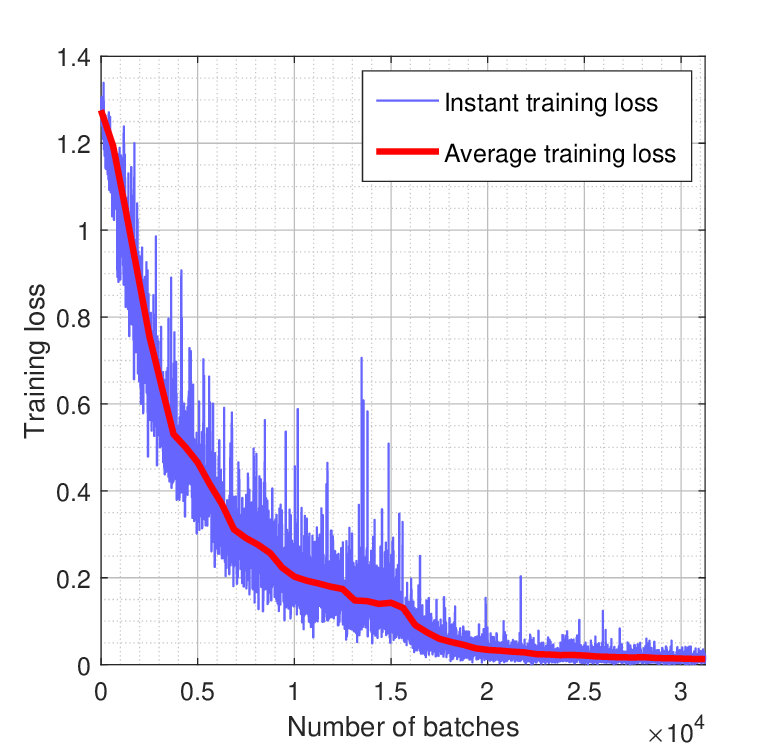}
\caption{Training loss versus the number of batches.}\label{fig4}
\end{figure}

Fig. \,\ref{fig4} displays the training loss along with the increasing number of batches in the case of  $\eta=0.001$, ${N_{{\rm{stp}}}} = 5$  and ${N_{{\rm{epo}}}} = 10$. Since each batch is comprised of $16$ data samples in the training stage, the number of batches for one epoch can be computed as  $3.125 \times {10^3}$, and there are total $3.125 \times {10^4}$  batches for $10$ epochs as shown in Fig. \ref{fig4}. To better understand the training process, the instant training loss is presented along with the number of batches in Fig. \ref{fig4}. Although the value of the instant training loss continuously changes for a specific batch, the overall trend of the instant training loss keeps lowering. Moreover, the average training loss is also drawn to further illustrate the rough tendency of the instant training loss. Remarkably, the fluctuation amplitude of the instant training loss dwindles with the growing batches.

\begin{figure}[!t]
\centering
\includegraphics[width=6.8cm]{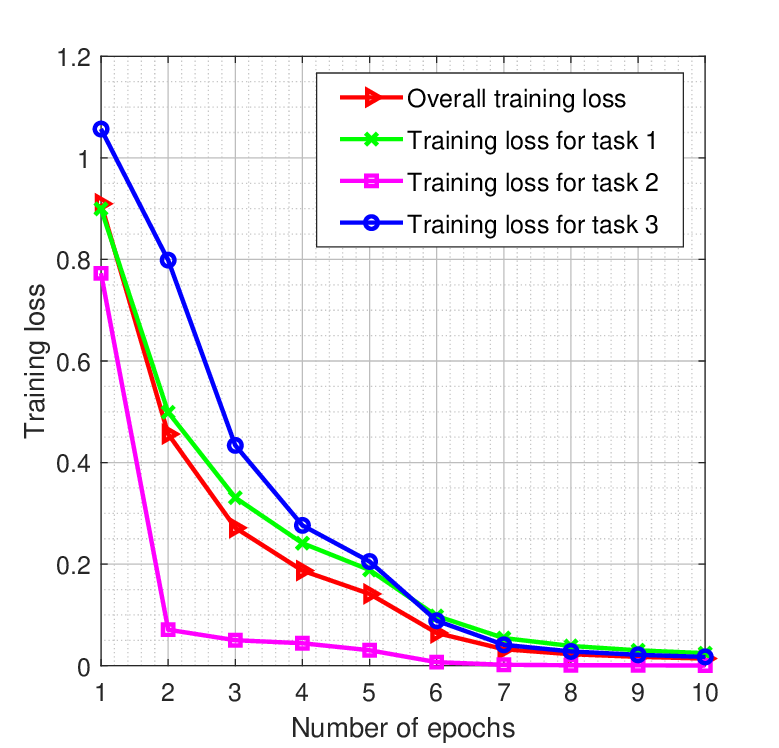}
\caption{Training loss versus the number of epochs.}\label{fig5}
\end{figure}

Fig. \,\ref{fig5} presents the training loss of three different kinds of tasks versus the number of epochs in the case of $\eta=0.001$, ${N_{{\rm{stp}}}} = 5$  and ${N_{{\rm{epo}}}} = 10$. From Fig. \,\ref{fig5}, we can observe that the training loss trends of three tasks are consistent with the overall training loss. Specifically, the training loss curve for task 2 converges faster than the rest tasks and achieves the lowest training loss. The training loss for all the tasks has already converged with ${N_{{\rm{epo}}}} = 10$. Hence, Fig. \,\ref{fig5} verifies that our proposed MTL-ABS model has the ability to solve multiple beam classifications simultaneously.

\begin{figure}[!t]
\centering
\includegraphics[width=6.8cm]{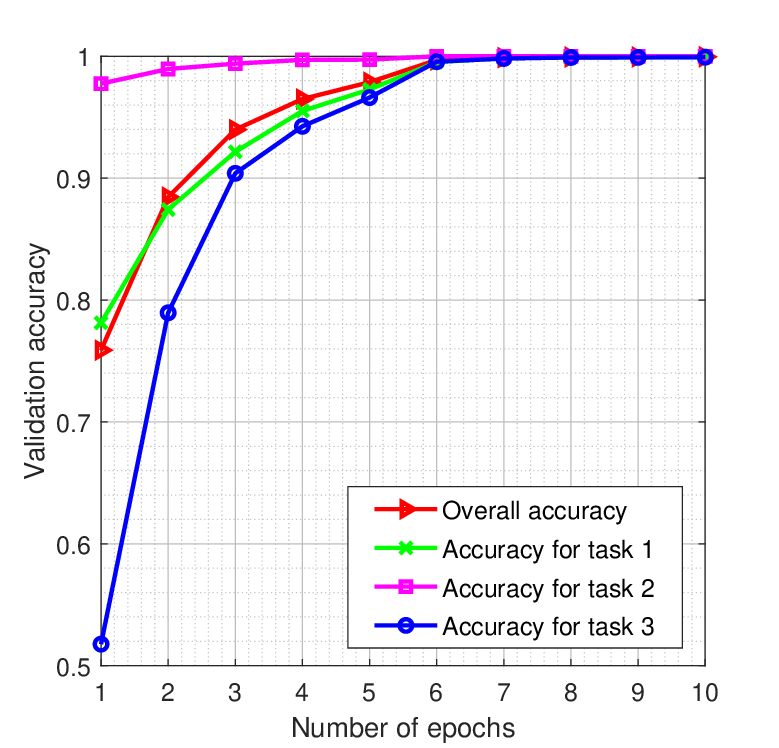}
\caption{Validation accuracy versus the number of epochs.}\label{fig6}
\end{figure}

Fig. \,\ref{fig6} shows the validation accuracy of multiple tasks along with the increasing number of epochs. From Fig. \ref{fig6}, the prediction accuracy of three tasks and the overall model becomes better when the number of epochs increases. At the beginning of the validation stage, task 2 possesses the best prediction accuracy while the accuracy for task 3 is relatively lower than other tasks. Remarkably, as long as the number of epochs is large enough, the validation accuracy is able to approach $100\%$. It is worth noting that these simulation results of the prediction performance in Fig. \,\ref{fig6} are consistent with the training results in Fig. \,\ref{fig5}, so the proposed MTL-ABS model has the strong generalization capability.

\subsection{Sum-rate and Complexity Comparisons}

\begin{figure}[!t]
\centering
    \includegraphics[width=6.8cm]{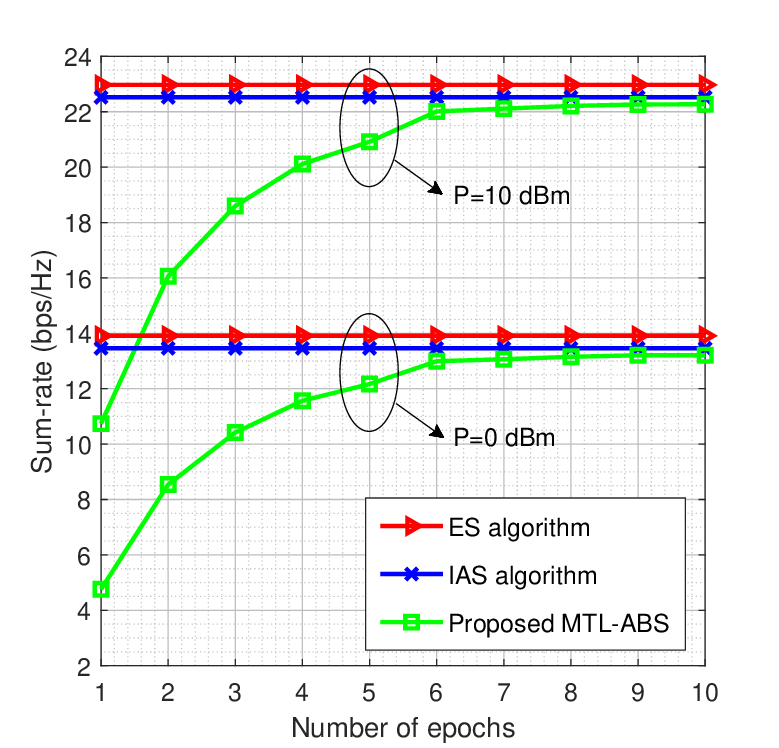}
\caption{Sum-rate comparisons versus the number of epochs.}\label{fig7}
\end{figure}

In this subsection, we explores the sum-rate comparisons of different analog beam selection schemes, e.g., ES, IAS and MTL-ABS. Fig. \ref{fig7} illustrates the achievable sum-rate of considered schemes versus the number of epochs with $P=0$ dBm and $P=10$ dBm. Here we select the parameters as $N_r=128$, $N_s=128$,  $M=128$, $M_s=2$, $N_t=16$ and $K=4$. From Fig. \ref{fig7} we note that the sum-rate performance of our proposed MTL-ABS scheme approaches the IAS algorithm and the optimal ES algorithm as the number of epochs increases, and the sum-rate gap among three schemes can be neglected (within 0.36 bps/Hz) in both $P=0$ dBm and $P=10$ dBm cases. To be specific, the sum-rate gap between the MTL-ABS scheme and the IAS algorithm is about 0.18 bps/Hz for $P=0$ dBm and 0.25 bps/Hz for $P=10$ dBm under the condition of ${N_{{\rm{epo}}}} = 10$. Consequently, with the sufficient number of epochs, our proposed MTL-ABS scheme is able to realize the near-optimal sum-rate performance compared with the heuristic search schemes.


\begin{figure}[!t]
\centering
\includegraphics[width=6.8cm]{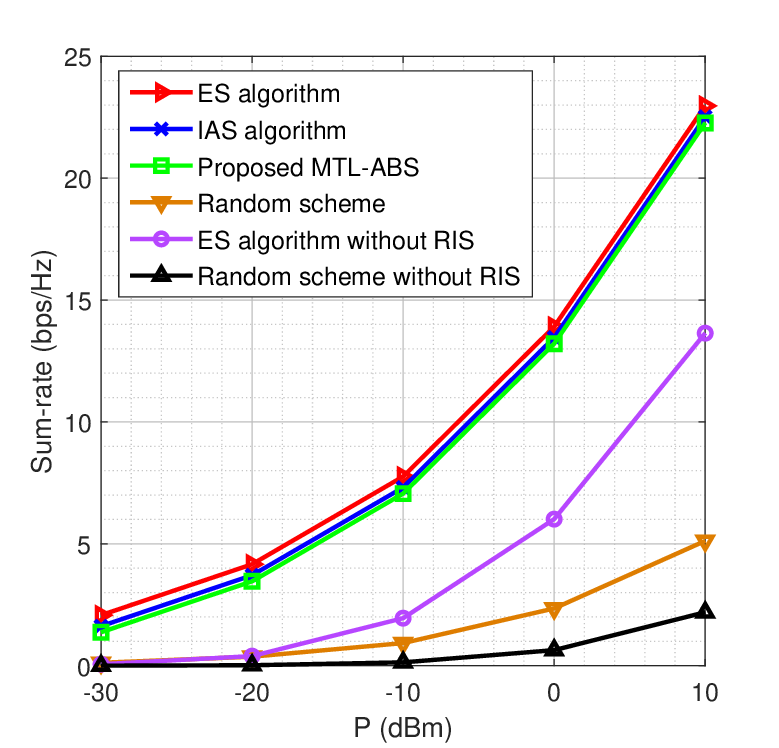}
\caption{Sum-rate comparisons versus transmit power $P$.}\label{fig8a}
\end{figure}

Fig. \,\ref{fig8a} investigates the sum-rate comparisons of considered beam selection schemes versus different transmit power in the case of $N_r=128$, $M=128$ and  $N_t=16$. From Fig. \ref{fig8a} we can observe that the sum-rate performance of the ES scheme and random scheme with RIS greatly outstrips the conventional ES scheme and random scheme without RIS, which reveals that the deployment of RIS definitely promotes the communication performance. Numerically, as shown in Fig. \ref{fig8a}, the RIS can bring extra 9.33 bps/Hz and 2.93 bps/Hz performance gains for the ES scheme and the random scheme at $P=10$ dBm. More importantly, the sum-rate curves of the IAS and MTL-ABS schemes are close to the optimal ES scheme, and the performance gaps (e.g., maximum 0.56 bps/Hz) among these three schemes can be neglected. 

\begin{figure}[!t]
\centering
\includegraphics[width=6.8cm]{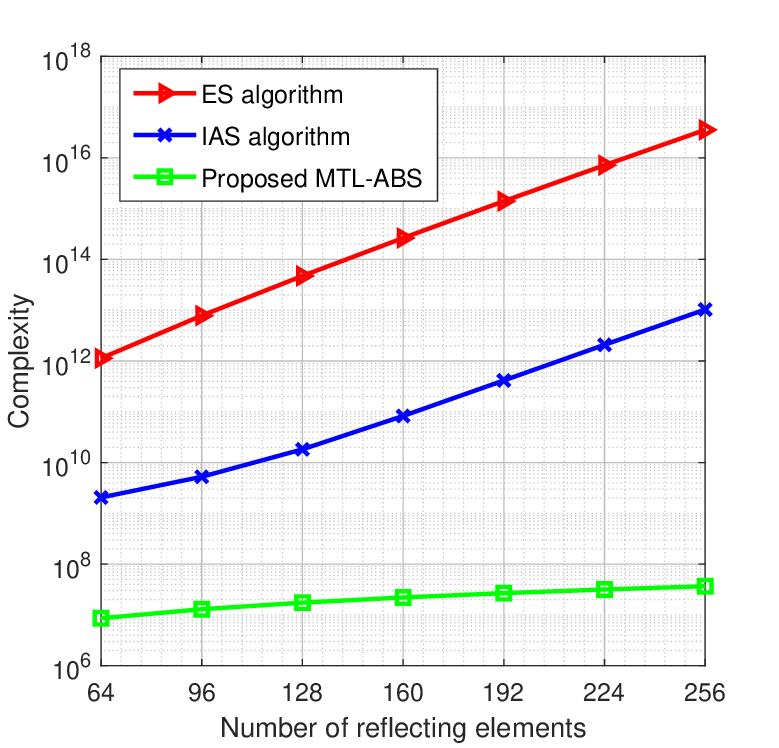}
\caption{Complexity comparisons of considered schemes versus the number of RIS elements.}\label{fig9}
\end{figure}

Fig. \ref{fig9} describes the complexity comparisons versus the number of RIS elements with $N_r=128$  and $N_t=16$, where the vertical coordinate is represented in the logarithmic coordinate. It is obvious that the proposed MTL-ABS scheme owns the lowest complexity, and the ES algorithm possesses the highest the complexity. Although the IAS algorithm substantially lessens the computation burden, it is still not applicable to real-time THz communications. In Fig. \ref{fig9}, the number of multiplications for the considered three schemes with $M=128$ can be counted as $1.18 \times {10^{13}}$, $7.94 \times {10^{9}}$ and $1.75 \times {10^{7}}$, respectively. Especially, the ES algorithm and the IAS algorithm endure the exponential growth of complexity as the number of RIS elements increases. Hence, the MTL-ABS scheme is able to meet the application scenarios with a large number of RIS elements.

\section{Conclusion}
In this paper, we handled the non-convex and cooperative beam selection problem with multiple classification tasks by evaluating the sum-rate metric. To restrict the complexity of explosive growth experienced by the ES and the IAS algorithms, we developed an efficient MTL-ABS framework to realize the multiple beam classification tasks in a low-complexity way. More importantly, the MTL-ABS framework can be extended to more complicated THz scenarios that will be encountered in 6G wireless networks.

\section*{Appendix A \\ Proof for Proposition 1}
To simplify the notation, we define that ${\bf{M}} = {{\bf{\bar { H}}}^H}{\bf{\bar { H}}}$, then according to the definition of inversion operation, ${{\bf{M}}^{ - 1}}$ can be expressed as
\begin{align}\label{Appendix01}
{{\bf{M}}^{ - 1}} = \frac{1}{{\det \left( {\bf{M}} \right)}}\left[ {\begin{array}{*{20}{c}}
{{{\bf{M}}_{1,1}}}&{{{\bf{M}}_{2,1}}}& \cdots &{{{\bf{M}}_{K,1}}}\\
{{{\bf{M}}_{1,2}}}&{{{\bf{M}}_{2,2}}}& \cdots &{{{\bf{M}}_{K,2}}}\\
 \vdots & \vdots & \ddots & \vdots \\
{{{\bf{M}}_{1,K}}}&{{{\bf{M}}_{2,K}}}& \cdots &{{{\bf{M}}_{K,K}}}
\end{array}} \right],
\end{align}
where ${{\bf{M}}_{i,j}}(\forall i,j )$  denotes the cofactor of $(i,j){\rm{th}}$  entry of ${{\bf{M}}}$. Hence, the complexity for calculating ${{\bf{M}}^{ - 1}}$  in a conventional way can be given by ${{\cal O}_{Con}}\left( {{K^2}{{\left( {K - 1} \right)}^3} + {K^3} + {K^2}} \right)$.

Then, according to (\ref{Appendix01}), we can further get that
\begin{align}\label{Appendix02}
{\left[ {{{\bf{M}}^{ - 1}}} \right]_{i,i}} = \frac{{{{\bf{M}}_{i,i}}}}{{\det \left( {\bf{M}} \right)}} = {\left( { - 1} \right)^{i + i}}\frac{{\det \left( {{{\bf{M}}_{\left\langle {i,i} \right\rangle }}} \right)}}{{\det \left( {\bf{M}} \right)}},
\end{align}
where ${{\bf{M}}_{\left\langle {i,i} \right\rangle }}$ indicates a minor of ${\bf{M}}$  by removing its $i$th  column and $i$th  row, which can be further formulated as
\begin{align}\label{Appendix03}
\det \left( {{{\bf{M}}_{\left\langle {i,i} \right\rangle }}} \right) & = \det \left( {{{\left( {{{{\bf{\bar {H}}}}^H}{\bf{\bar {H}}}} \right)}_{\left\langle {i,i} \right\rangle }}} \right) \notag \\
& = \det \left( {{{\left[ {\begin{array}{*{20}{c}}
{{\bf{\bar h}}_1^H{{{\bf{\bar h}}}_1}}&{{\bf{\bar h}}_1^H{{{\bf{\bar h}}}_2}}& \cdots &{{\bf{\bar h}}_1^H{{{\bf{\bar h}}}_K}}\\
{{\bf{\bar h}}_2^H{{{\bf{\bar h}}}_1}}&{{\bf{\bar h}}_2^H{{{\bf{\bar h}}}_2}}& \cdots &{{\bf{\bar h}}_2^H{{{\bf{\bar h}}}_K}}\\
 \vdots & \vdots & \ddots & \vdots \\
{{\bf{\bar h}}_K^H{{{\bf{\bar h}}}_1}}&{{\bf{\bar h}}_K^H{{{\bf{\bar h}}}_2}}& \cdots &{{\bf{\bar h}}_K^H{{{\bf{\bar h}}}_K}}
\end{array}} \right]}_{\left\langle {i,i} \right\rangle }}} \right) \notag \\
& = \det \left( {{\bf{\bar {H}}}_{\left\langle {i, \sim } \right\rangle }^H{{{\bf{\bar {H}}}}_{\left\langle { \sim ,i} \right\rangle }}} \right).
\end{align}

By considering (\ref{Appendix02}) and (\ref{Appendix03}), we can get the result of \emph{Proposition 1}. In addition, the complexity of calculating (\ref{IAS01}) can be given by ${{\cal O}_{New}}\left( {2{K^3} + 3{K^2} + 3K} \right)$. Hence, the proposed sum-rate metric in (\ref{IAS02}) greatly reduces the complexity, and \emph{Proposition 1}  has been completely proven.

\section*{Appendix B \\ Proof for Proposition 2}
According to \cite{BWC_02}, problem (\ref{BWC01}) is able to be defined as a more general form, which can be given by
\begin{align}\label{BWC06}
{F}\left( {{\bf{G}},{\bf{Q}}} \right) = {f_0}\left( {{\bf{G}},{\bf{Q}}} \right) + {f_S}\left( {\bf{G}} \right) + {f_T}\left( {\bf{Q}} \right),
\end{align}
where the penalty terms ${f_S}\left( {\bf{G}} \right)$ and ${f_T}\left( {\bf{Q}} \right)$  are differentiable convex functions. In addition, the continuously differentiable function ${f_0}\left( {{\bf{G}},{\bf{Q}}} \right)$  is the bi-convex with ${\bf{G}}$  and ${\bf{Q}}$. Then we define the partial derivatives ${\nabla _{\bf{G}}}{f_1}\left( {{\bf{G}},{\bf{Q}}} \right)$  and ${\nabla _{\bf{Q}}}{f_1}\left( {{\bf{G}},{\bf{Q}}} \right)$ that correspond to ${\bf{G}}$  and ${\bf{Q}}$, respectively. Under these notations, ${\nabla _{\bf{G}}}{f_0}\left( {{\bf{G}},{\bf{Q}}} \right)$  holds the following condition that
\begin{align}\label{BWC07}
&\left\| {{\nabla _{\bf{G}}}{f_0}\left( {{\bf{G}} + {{\bf{\Delta }}_{\bf{G}}},{\bf{Q}}} \right) - {\nabla _{\bf{G}}}{f_0}\left( {{\bf{G}},{\bf{Q}}} \right)} \right\|_F \notag \\
& = \left\| {\sum\limits_{u = 1}^U {{{\bf{X}}^T}{\bf{X}}{{\bf{\Delta }}_{\bf{G}}}{{\bf{F}}_S}{{\bf{Q}}_u}{{\bf{F}}_u}{\bf{F}}_u^T{\bf{Q}}_u^T} {\bf{F}}_S^T} \right\|_F \notag \\
& = \left\| {\sum\limits_{u = 1}^U {{{\left( {{{\bf{F}}_S}{{\bf{Q}}_u}{{\bf{F}}_u}{\bf{F}}_u^T{\bf{Q}}_u^T{\bf{F}}_S^T} \right)}^T} \otimes \left( {{{\bf{X}}^T}{\bf{X}}} \right){\rm {vec}}\left( {{{\bf{\Delta }}_{\bf{G}}}} \right)} } \right\|_F \notag \\
& \le \left\| {\sum\limits_{u = 1}^U {{{\left( {{{\bf{F}}_S}{{\bf{Q}}_u}{{\bf{F}}_u}{\bf{F}}_u^T{\bf{Q}}_u^T{\bf{F}}_S^T} \right)}^T} \otimes \left( {{{\bf{X}}^T}{\bf{X}}} \right)} } \right\|_F \left\| {{{\bf{\Delta }}_{\bf{G}}}} \right\|_F \notag \\
&\le {L_{\bf{G}}}\left\| {{{\bf{\Delta }}_{\bf{G}}}} \right\|_F,
\end{align}
where ${{\bf{\Delta }}_{\bf{G}}} \in {{\mathbb C}^{r \times n}}$  and ${L_{\bf{G}}}$  denotes the Lipschitz constant for  ${\bf{G}}$. This derived property shows that the gradient of ${f_0}\left( {{\bf{G}},{\bf{Q}}} \right)$  is Lipschitz continuous with regard to ${\bf{G}}$. In the case of optimizing ${\bf{Q}}$, ${\nabla _{\bf{Q}}}{f_0}\left( {{\bf{G}},{\bf{Q}}} \right)$  also possesses the identical Lipschitz property as follow
\begin{align}\label{BWC08}
& \left\| {{\nabla _{\bf{Q}}}{f_0}\left( {{\bf{G}} + {{\bf{\Delta }}_{\bf{Q}}},{\bf{Q}}} \right) - {\nabla _{\bf{Q}}}{f_0}\left( {{\bf{G}},{\bf{Q}}} \right)} \right\|_F \notag \\
& = \left\| {\sum\limits_{u = 1}^U {{\bf{F}}_S^T{{\bf{G}}^T}{{\bf{X}}^T}{\bf{XG}}{{\bf{F}}_S}{{\bf{\Delta }}_{{\bf{Q}},u}}{{\bf{F}}_u}{\bf{F}}_u^T} } \right\|_F \notag \\
& = \left\| {\sum\limits_{u = 1}^U {\left( {{{\bf{F}}_u}{\bf{F}}_u^T} \right) \otimes \left( {{\bf{F}}_S^T{{\bf{G}}^T}{{\bf{X}}^T}{\bf{XG}}{{\bf{F}}_S}} \right){\rm {vec}}\left( {{{\bf{\Delta }}_{{\bf{Q}},u}}} \right)} } \right\|_F \notag \\
& = \left\| {\left[ {{{\bf{\Pi }}_1},{{\bf{\Pi }}_2}, \cdots ,{{\bf{\Pi }}_U}} \right]\left[ {\begin{array}{*{20}{c}}
{{{\bf{\Delta }}_{{\bf{Q}},1}}}\\
{{{\bf{\Delta }}_{{\bf{Q}},2}}}\\
 \vdots \\ {{{\bf{\Delta }}_{{\bf{Q}},U}}} \end{array}} \right]} \right\|_F \notag \\
& \le \left\| {\left[ {{{\bf{\Pi }}_1},{{\bf{\Pi }}_2}, \cdots ,{{\bf{\Pi }}_U}} \right]} \right\|_F \left\| {{{\bf{\Delta }}_{\bf{Q}}}} \right\|_F \notag \\
& \le {L_{\bf{Q}}}\left\| {{{\bf{\Delta }}_{\bf{Q}}}} \right\|_F,
\end{align}
where ${{\bf{\Delta }}_{\bf{Q}}} = {[{\bf{\Delta }}_{{\bf{Q}},1}^T,{\bf{\Delta }}_{{\bf{Q}},2}^T, \cdots ,{\bf{\Delta }}_{{\bf{Q}},U}^T]^T} \in {{\mathbb C}^{q \times ({d_1} +  \cdots  + {d_U})}}$, ${{\bf{\Pi }}_{u \in {\cal U}}} = \left( {{{\bf{F}}_u}{\bf{F}}_u^T} \right) \otimes \left( {{\bf{F}}_S^T{{\bf{G}}^T}{{\bf{X}}^T}{\bf{XG}}{{\bf{F}}_S}} \right)$, and ${L_{\bf{Q}}}$  is the Lipschitz constant for ${\bf{Q}}$, respectively. Similar to ${\nabla _{\bf{G}}}{f_0}\left( {{\bf{G}},{\bf{Q}}} \right)$, ${\nabla _{\bf{Q}}}{f_0}\left( {{\bf{G}},{\bf{Q}}} \right)$  also satisfies the Lipschitz continuous condition. Hence, by utilizing Lemma 3.1 in \cite{BWC_02}, the network weight sequence $\{ {{\bf{G}}^i},{{\bf{Q}}^i}\}$  generated by our proposed blockwise MTL-ABS framework proves to be a stationary point  $\{ {{\bf{G}}^{\star}},{{\bf{Q}}^{\star}}\}$  under the case of $i \to \infty$. Hence, the proof for \emph{Proposition 2} is completed.

\end{document}